\documentclass[aps,prap,reprint,groupedaddress]{revtex4-2}
\usepackage{bm}
\usepackage{amsmath}
\usepackage{amssymb}
\usepackage{graphicx}
\usepackage{color}
\begin{document}

\title{Large angle precession of magnetization maintained by a microwave
  voltage}
\author{Hiroshi Imamura}
\email[]{h-imamura@aist.go.jp}
\altaffiliation{National Institute of Advanced Industrial Science and
  Technology (AIST), Research Center for Emerging Computing Technologies, Tsukuba, Ibaraki 305-8568, Japan}
\author{Rie Matsumoto}
\email[]{rie-matsumoto@aist.go.jp}
\altaffiliation{National Institute of Advanced Industrial Science and
  Technology (AIST), Research Center for Emerging Computing Technologies, Tsukuba, Ibaraki 305-8568, Japan}

\begin{abstract}
Effects of a microwave voltage on magnetization precession are
analyzed based on a macrospin model. The microwave voltage induces the
oscillating anisotropy field through the voltage controlled magnetic
anisotropy (VCMA) effect, and then stimulates the magnetization. The
large angle precession is maintained if the magnetization synchronizes
with the microwave voltage.  The effective equations of motion of the
magnetization with an oscillating anisotropy field are
derived, and the mechanism of the synchronization is clarified by
analyzing the derived equations of motion. The conditions of the
angular frequency detuning and the amplitude of the oscillating
anisotropy field for synchronization are obtained. The results are
useful for development of the VCMA-based energy-efficient spintronics
devices using magnetization precession such as a VCMA-based
magnetoresistive random access memory and a nano-scale microwave
magnetic field generator.
\end{abstract}

\maketitle

\section{introduction}
Magnetic anisotropy (MA) is a key property of a ferromagnet, which
stabilizes the direction of magnetization even at room temperature
\cite{Chikazumi,Landau8}.  In a magnetoresistive random access memory
(MRAM) the information is stored as the direction of magnetization,
e.g. up or down.  The height of the energy barrier between the up and
down states is proportional to the MA constant, and the retention time
of the information is an exponential increasing function of the MA
constant \cite{Brown1963}.  The current standard writing scheme of
MRAM utilizes the spin transfer torque (STT)
\cite{Slonczewski1996,Berger1996,Stiles2005} because of low power
consumption and high integration density compared with the magnetic
field switching. The STT acts as the negative damping torque and
excites the magnetization over the energy barrier.

The discovery of voltage controlled magnetic anisotropy (VCMA) effect
\cite{Weisheit2007,Duan2008,Tsujikawa2009,Nakamura2009,Maruyama2009,Nozaki2010,Endo2010}
provides a more energy-efficient writing scheme of MRAM. In the
VCMA-based switching
\cite{Shiota2012,Kanai2012,Grezes2016,Kanai2016,Matsumoto2018,Yamamoto2018,Imamura2019,Matsumoto2019,Yamamoto2019,Yamamoto2020,Matsumoto2020},
application of the voltage to the MRAM cell eliminates the MA of the
free layer (FL) and induces the precession of magnetization around the
external magnetic field.  The switching completes if the voltage is
turned off at a half period of precession.  The power consumption of
the VCMA-based switching is much smaller than that of the STT-based
switching because of little Joule heating \cite{Grezes2016,Kanai2016}.

  Parametric excitation (PE) is a nonlinear phenomenon induced
  by periodic modulation of a parameter in equations of
  motion and, which has been studied in many areas of physics and engineering
  \cite{doi:10.1002/9783527617586.ch5}.
  In magnetic materials PE of magnetization is induced by applying a
  microwave field \cite{Bloembergen1952,Bloembergen1954,Anderson1955,Urazhdin2010,Urazhdin2010a,Ulrichs2011,Martin2011,Edwards2012,Zhu2008,Lu2013,Zhou2017},
  microwave current \cite{Cui2008,Durrenfeld2014,Montoya2019}, or
  microwave voltage \cite{Verba2014,Verba2016,Verba2017,Chen2017,Yamamoto2020Nano}.
  The PE of the spin waves by microwave voltage in ferromagnetic films
  \cite{Verba2014,Verba2016,Verba2017} attracts
  much attention as a key element of low power consumption magnonic
  devices \cite{Rana2019}.

Very recently Yamamoto et al. studied the effect of PE on the
VCMA-based switching in an MRAM cell and found that
the oscillation amplitude of the switching probability does not decay if
the magnetization precesses in synchronization with the applied
oscillating voltage \cite{Yamamoto2020Nano}. They also performed the
numerical simulations based on the macrospin model and reproduced the
experimental results very well.
Although the results are qualitatively and intuitively explained by
the concept of PE, more detailed theoretical analysis on the mechanism as
well as the conditions for the PE in this system is necessary for
practical applications.
Since the experimental situation is quite
different from the PE of spin waves in ferromagnetic films, the theory
developed by Verba et al. \cite{Verba2014,Verba2016,Verba2017} is not
directly applicable for the analysis of the results in
Ref. \cite{Yamamoto2020Nano}.

In this paper, we analyze the effects of a microwave voltage on
magnetization precession using the macrospin model following the standard analysis of PE \cite{doi:10.1002/9783527617586.ch5}.
It is shown that the large angle precession is maintained if the
magnetization synchronizes with the microwave voltage.  The mechanism of the
synchronization is clarified, and the conditions of parameters for 
synchronization are obtained by analyzing the effective equations of
motion of the magnetization under the microwave voltage.

  The results are useful for reducing the write error rate of the VCMA-based MRAM by using a microwave voltage pulse.  The large angle precession of magnetization maintained by a microwave voltage can be applied as a low-power nano-scale microwave magnetic field generator.

%============================================================
% Model and Method
%============================================================
\section{model and method}

\begin{figure}[t]
\centerline{\includegraphics[width=\columnwidth]{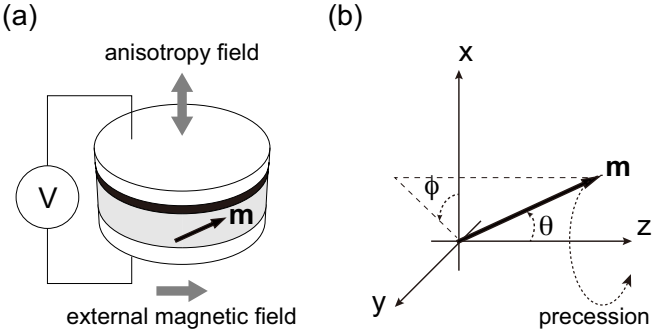}}
\caption{
  \label{fig:model}
  (a) Schematic illustration of the FL in the VCMA-based MRAM.
  The FL shown in gray is a thin magnetic layer with
  a perpendicular MA.
  $\bm{m}$ represents the magnetization unit vector.
  The insulating layer is shown in black. The other layers such as
  the reference layer and electrodes are
  shown in white. The external magnetic field is applied to the
  in-plane direction. The anisotropy field is modified by application
  of a voltage, $V$.
  (b) Definitions of the Cartesian coordinates, and the polar angle, $\theta$,
  and azimuthal angle, $\phi$, of $\bm{m}$. 
  The perepndicular anisotropy field is parallel to the $x$ axis. The external field
  is applied to the positive $z$ direction.
  The direction of  magnetization precession is indicated by the
  dotted circular arrow.
  }
\end{figure}

The system we consider is schematically shown in
Fig. \ref{fig:model}(a).  The FL of a VCMA-based MRAM with
perpendicular MA is shown in gray. Application of a voltage, $V$,
modifies the electron state at the interface between the FL and the
insulating layer shown in black and changes the MA of the FL through
the VCMA effect. The white cylinders represent the other layers such
as the reference layer and electrodes.  The definition of the
coordinate system is shown in Fig. \ref{fig:model}(b).  In terms of
the polar angle, $\theta$, and the azimuthal angle, $\phi$, the
magnetization unit vector, $\bm{m}$, is expressed as $\bm{m}$ =
($m_x$,$m_y$,$m_z$) =
($\sin\theta\cos\phi$,$\sin\theta\sin\phi$,$\cos\theta$).  The
external magnetic field, $H_{\rm ext}$, is applied to the positive $z$
direction and the perpendicular anisotropy field is parallel to the $x$ axis.  The
voltage is assumed to be the sum of the static and oscillating
components expressed as $ V = V_{\rm st} + V_{\rm osc}$.  The static
component, $V_{\rm st}$, is assumed to be large enough to eliminate
the static MA of the FL. The oscillating component, $V_{\rm osc}$,
generates the oscillating anisotropy field and stimulates the
magnetization.

In order to simplify the notation, the dimensionless forms of the
magnetic field and time are introduced.  The external field, $H_{\rm
  ext}$, is taken as the unit of the magnetic field.  The unit of time
is set $(1+\alpha^{2})/(\gamma H_{\rm ext})$, where $\alpha$ is the
Gilbert damping constant and $\gamma$ is the gyromagnetic ratio.

The oscillating anisotropy field induced by $V_{\rm osc}$ is given by
\begin{align}
  h_{\rm k} \cos(\omega\tau) \sin\theta\cos\phi,
\end{align}
where $h_{\rm k}$ is the amplitude of the oscillating anisotropy
field, and $\omega$ and $\tau$ are the angular frequency of $V_{\rm
  osc}$ and time in the dimensionless unit, respectively.  The
dynamics of $\bm{m}$ is obtained by solving the following
Landau-Lifshitz-Gilbert (LLG) equation,
\begin{align}
  \label{eq:LLGtheta}
  \dot{\theta} =
  & -\alpha \sin\theta - h_{\rm k}\cos(\omega\tau)
  \sin\theta \cos\phi, \notag\\
  & \times(\sin\phi - \alpha \cos\theta \cos\phi) \\
  \label{eq:LLGphi}
  \dot{\phi} =
  & 1 - h_{\rm k}\cos(\omega\tau) \cos\phi
  (\cos\theta \cos\phi + \alpha \sin\theta),
\end{align}
where the dot represents the derivative in terms of $\tau$.  The
values of $\alpha$ and $h_{\rm k}$ are assumed to be much smaller than
unity. For numerical simulations and exemplification of the analytical
results we assume $\alpha = 0.01$ and $h_{\rm k} = 0.05$ unless
otherwise mentioned.  In numerical simulations the 4th order
Runge-Kutta method is employed to solve the LLG equation.
  Before application of the voltage the magnetization is aligned
perpendicular to the plane because the FL has a static perpendicular
MA.
The initial state is assumed to be $m_{x} = 1$, i.e. $\theta = \pi/2$
and $\phi = 0$.

\section{Results and Discussions}
%========================================
% Fig. 2
%========================================
\begin{figure}[t]
  \centerline{
    \includegraphics[width=\columnwidth]{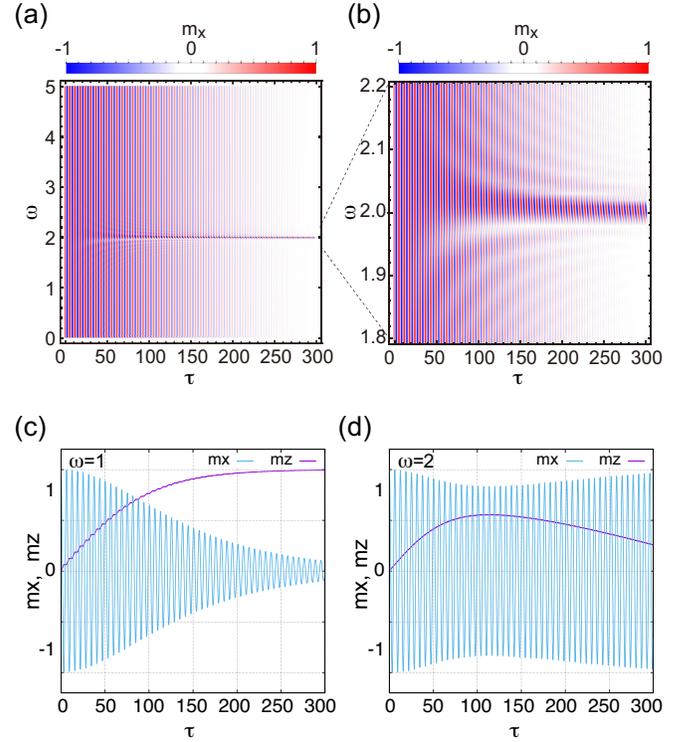}}
\caption{
  \label{fig:t_f_mz_mxyz}
  (Color online)
  (a)
  A color density plot of $m_{x}$ on the $\omega$-$\tau$ plane.
  Around $\omega = 2$ the large angle precession of magnetization is
  maintained by the microwave voltage.
  (b)
  The enlarged view of panel (a) for $1.8 \le \omega \le 2.2$.
  The synchronization region is about $1.98 \lessapprox \omega
  \lessapprox 2.02$ at $\tau = 300$.
  (c) The dynamics of $m_{x}$ and $m_{z}$ for
  the non-synchronized state at $\omega = 1$.
  The precession amplitude decays monotonically.
  (d) The same plot as (c) for the synchronized state at $\omega = 2$. The
  precession amplitude increases with increase of $\tau$ for $\tau
  \gtrapprox 100$.
    }
\end{figure}

Figure \ref{fig:t_f_mz_mxyz}(a) is the color density plot of $m_{x}$ on
$\tau$-$\omega$ plane, which can be observed
%
% experimentally
%
owing to the magnetoresistance effect. 
The large amplitude of oscillation is maintained around twice the
natural angular frequency of precession, i.e. $\omega = 2$. 
In Ref. \citenum{Yamamoto2020Nano},
the similar plots were obtained for
the switching probabilities.
The enlarged view for $1.8 \le \omega \le 2.2$ is shown in
Fig. \ref{fig:t_f_mz_mxyz}(b).
The region of angular frequency where the large amplitude of oscillation is
maintained is $1.98 \lessapprox \omega \lessapprox 2.02$.
Outside of this synchronization region the oscillation amplitude shows
a monotonic decay.

The dynamics of $m_{x}$ and $m_{z}$ at $\omega = 1$ is plotted in
Fig. \ref{fig:t_f_mz_mxyz}(c). $m_{z}$ increases with increase of
$\tau$ and approaches to unity. $m_{x}$ shows the damped oscillation
with a period of $2 \pi$. These results agree well with the exact
solutions for $h_{\rm k} = 0$ \cite{Landau8}, which implies that the
$V_{\rm osc}$ with $\omega = 1$ has little effect on the magnetization
dynamics.

The same plot at $\omega = 2$ are shown in Fig. \ref{fig:t_f_mz_mxyz}(d).
Contrary to the results at $\omega = 1$, the oscillation amplitude does
not show a monotonic decay.
Although the oscillation amplitude decays for $\tau
\lessapprox 100$, it takes the minimum value at $\tau \approx 100$ and then
increases with increase of $\tau$.
The period of oscillation is $2 \pi$, i.e. $\bm{m}$
oscillates with the natural angular frequency of unity instead of the
angular frequency of $V_{\rm osc}$, $\omega = 2$.
The fact that the precession amplitude of $\bm{m}$ is strongly
enhanced by the external periodic force with twice the natural
angular frequency implies that this phenomena is closely related to the
parametric excitation
\cite{Bloembergen1952,Bloembergen1954,Anderson1955,Urazhdin2010a,Ulrichs2011,Martin2011,Edwards2012,Durrenfeld2014,Chen2017b}.

%========================================
% Fig 3
%========================================
\begin{figure}[t]
  \centerline{
    \includegraphics[width=\columnwidth]{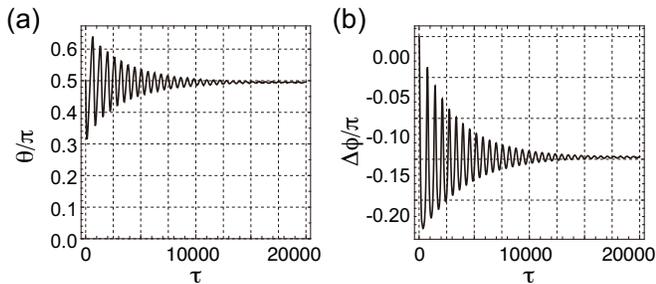}}
\caption{
  \label{fig:qft}
  (a) The dynamics $\theta$ at $\omega = 2$ for $0\le\tau\le$ 20000.
  It oscillates with a period much longer
  than $2\pi$ and converges to about $\pi/2$ for large $\tau$.
  (b) The same plot as (a) for $\Delta\phi$. 
  It oscillates with the same period as $\theta$ and
  converges to about $-0.148 \pi$ for large $\tau$.
  }
\end{figure}

Figure \ref{fig:qft}(a) shows the dynamics of 
the polar angle, $\theta$, at $\omega = 2$ for a long time duration
of $\tau \le 20000$. 
Figure \ref{fig:qft}(b) is the same plot for the phase shift from the
free precession at the angular frequency of $\omega/2$, which is defined as
\begin{align}
  \label{eq:phase}
  \Delta\phi = \phi-\frac{\omega}{2}\tau.
\end{align}
The polar angle is a measure of the oscillation amplitude because
$m_{x} = \sin\theta$. The azimuthal angle, $\phi$, represents the
oscillation phase, and the phase shift, $\Delta\phi$, is closely
related to the energy absorption.
The polar angle shows a slow oscillation with a period much longer
than $2\pi$ and converges to about $\pi/2$ for large $\tau$.
The phase shift also shows the slow oscillation with the same period as
the polar angle and converges to about $-0.148 \pi$ for large $\tau$.

The equation of motion for the slow dynamics of the polar angle is derived
as follows. Since both $\alpha$ and $h_{\rm k}$ are assumed to be much smaller
than unity, the terms with $\alpha h_{\rm k}$ can be neglected in the
LLG equation. Then Eq. \eqref{eq:LLGtheta} becomes
\begin{align}
  \dot{\theta} = -\alpha \sin\theta - h_{\rm k}\cos(\omega\tau) \sin\theta
  \cos\phi\sin\phi.
  \label{eq:LLGtheta2}
\end{align}
The second term on the right hand side depends on $\phi$ as
$\cos\phi\sin\phi$, where $\cos\phi$ comes from $m_{x}$, and $\sin\phi$
is the projection coefficient of the anisotropy field torque to the
direction of $\dot{\theta}$. Since $\cos\phi\sin\phi =
\sin(2\phi)/2$, $\omega$ should be twice the angular frequency of magnetization
precession to realize synchronization.

Substituting $\phi = \omega\tau/2 + \Delta\phi$ into
Eq. \eqref{eq:LLGtheta2} and applying the trigonometric identities we
obtain
\begin{align}
  \dot{\theta} = -\sin\theta \left\{
  \alpha
  +\frac{h_{\rm k}}{4}\left[
  \sin(2\Delta\phi) + \sin(2\omega\tau + 2\Delta\phi)
   \right]
  \right\}.
\end{align}
Since we are interested in the slow dynamics of $\theta$ and
$\Delta\phi$ we average out the fast oscillating term with
$\sin(2\omega\tau + 2\Delta\phi)$.
Finally the equation of motion for the slow dynamics of $\theta$ is
obtained as
\begin{align}
  \label{eq:dqdt}
  \dot{\theta} =
  & -\sin\theta \left[
    \alpha
    +\frac{h_{\rm k}}{4}
    \sin(2\Delta\phi) 
  \right].
\end{align}
Introducing the effective damping coefficient defined as
\begin{align}
  \label{eq:alpha_prime}
  \alpha^{\prime}
  =
  \alpha
  +\frac{h_{\rm k}}{4}
  \sin(2\Delta\phi),
\end{align}
Eq. \eqref{eq:dqdt} is expressed as $\dot{\theta} =
-\alpha^{\prime}\sin\theta$, which is the same form as Eq. \eqref{eq:LLGtheta}
with $h_{\rm k} = 0$. Assuming that  $\dot{\theta} = 0$ and
$\sin\theta\neq 0$ in the limit of $\tau\to\infty$,
the convergence value of the phase shift, $\Delta\phi_{\rm c}$, should be
adjusted to satisfy
$\sin(2\Delta\phi_{\rm c}) = -4\alpha/h_{\rm k}$.
There are two kinds of solutions for this equation.
One is
\begin{align}
  \Delta\phi_{\rm c}^{(-)}
  = -\frac{1}{2}\arcsin\left(\frac{4\alpha}{h_{\rm k}}\right),
\end{align}
and the other is
\begin{align}
  \Delta\phi_{\rm c}^{(+)} 
  = -\frac{\pi}{2}+\frac{1}{2}\arcsin\left(\frac{4\alpha}{h_{\rm k}}\right).
\end{align}
For the parameters we assumed, i.e. $\alpha = 0.01$ and $h_{\rm
  k} = 0.05$, $\Delta\phi_{\rm c}^{(-)} = -0.148 \pi$ which is identical
to the numerical results shown in Fig. \ref{fig:qft}(b).

The equation of motion for the slow dynamics of $\Delta\phi$ is
obtained in a similar manner.
Substituting  $\phi = \omega\tau/2 + \Delta\phi$ into
Eq. \eqref{eq:LLGphi}, neglecting the term with $\alpha h_{\rm k}$,
and averaging out the
fast oscillating term, we obtain
\begin{align}
  \label{eq:dfdt}
  \dot{\Delta\phi} = 1-\frac{\omega}{2} - \frac{h_{\rm k}}{4}\cos\theta
  \cos (2\Delta\phi).
\end{align}
Equation \eqref{eq:dfdt} is reduced to $\dot{\Delta\phi} = -
\frac{h_{\rm k}}{4}\cos\theta \cos (2\Delta\phi)$ at $\omega = 2$.
Assuming that  $\cos(2\Delta\phi)>0$, i.e. $|\Delta\phi| < \pi/2$
, the sign of $\dot{\Delta\phi}$ is determined by the sign of $\cos\theta$.
$\Delta\phi$ increases (decreases) with increase of $\tau$ if $\theta >
\pi/2 $ ($\theta < \pi/2)$.

\begin{figure}
  \centerline{
    \includegraphics[width=\columnwidth]{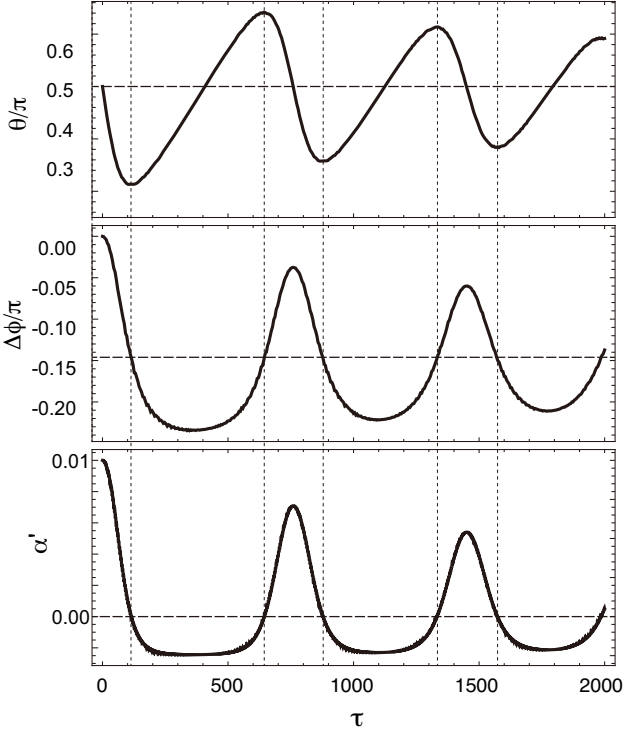}}
\caption{
  \label{fig:qf-Ea-t}
  The dynamics of $\theta$ (top), $\Delta\phi$ 
  (middle), and $\alpha^{\prime}$ (bottom)  at $\omega = 2$ for
  $0\le\tau\le$ 2000.
  The horizontal dashed lines in the top, middle, and bottom panels
  represent the values of $1/2$, $\Delta\phi_{\rm
    c}^{(-)} / \pi$,  and zero, respectively.
  The vertical dotted lines represent the values of $\tau$ at which
  $\Delta\phi$ takes the convergence value of $\Delta\phi_{\rm c}^{(-)}$.
}
\end{figure}

To understand the mechanism of the synchronization we analyze
the dynamics of $\theta$, $\Delta\phi$, and $\alpha^{\prime}$ in the
first two and a half period of oscillation. Figure \ref{fig:qf-Ea-t}
shows the dynamics of $\theta$ (top), $\Delta\phi$ (middle), and
$\alpha^{\prime}$ (bottom) at $\omega = 2$ for $0\le\tau\le 2000$.
The horizontal dashed lines
in the top, middle, and bottom panels represent the values of $1/2$,
$\Delta\phi_{\rm c}^{(-)} / \pi$, and zero, respectively.
The vertical dotted lines represent the values of $\tau$ at which
$\Delta\phi$ takes the convergence value of $\Delta\phi_{\rm c}^{(-)}$.

%========================================
% Fig 4
%========================================

At the beginning of the dynamics, $\theta$ decreases with increase of
$\tau$ because $\alpha^{\prime} \simeq \alpha$ ($>$0). $\Delta\phi$ also
decreases with increase of $\tau$ because $\theta < \pi/2$.
The decrease of $\Delta\phi$ induces the reduction of $\alpha^{\prime}$
following Eq. \eqref{eq:alpha_prime}.
When $\Delta\phi$ crosses the horizontal dashed line, i.e.
$\Delta\phi = \Delta\phi_{\rm c}^{(-)}$, $\theta$ takes the minimum value and
then starts to increase with increase of $\tau$ because $\alpha^{\prime}$
becomes negative. However $\Delta\phi$ and $\alpha^{\prime}$ 
decrease with increase of $\tau$ until $\theta$ exceeds $\pi/2$.
At $\tau \approx 400$, $\theta$ crosses the horizontal dashed line at
$\theta = \pi/2$. Then both $\Delta \phi$ and
$\alpha^{\prime}$ take the minimum values and start to increase
with increase of $\tau$.
Repeating the above procedure $\theta$, $\Delta\phi$, and $\alpha$
oscillate with the same period as each other and converge to the value of $\pi/2$, $\Delta\phi_{\rm c}^{(-)}$, and zero, respectively.

Let us move on to the analysis of the effect of angular frequency detuning on
synchronization. The angular frequency detuning is defined as the difference between the
angular frequency of $V_{\rm osc}$ and twice the natural angular
frequency of magnetization precession, i.e. $\nu = \omega - 2$. As shown in the
Fig. \ref{fig:t_f_mz_mxyz}(b) $\bm{m}$ synchronizes with the
$V_{\rm osc}$ within the range of 
$-0.02 \lessapprox \nu \lessapprox 0.02$ at $\tau = 300$.
The synchronization region of $\nu$ in the limit of $\tau\to\infty$ is
obtained as follows.
Substituting $\omega = 2 + \nu$ into Eq. \eqref{eq:dfdt} we obtain
\begin{align}
  \dot{\Delta\phi}
  = -\frac{\nu}{2}
  -\frac{h_{\rm k}}{4}\cos\theta \cos (2\Delta\phi).
\end{align}
Assuming that $\lim_{\tau\to\infty}\dot{\Delta\phi} = 0$ and
$\lim_{\tau\to\infty}\Delta\phi = \Delta\phi_{\rm c}^{(-)}$ the
convergence value of $\theta$ is obtained by
solving
\begin{align}
   -\frac{\nu}{2}
  -\frac{h_{\rm k}}{4}\cos\theta_{\rm c} \cos (2\Delta\phi_{\rm
  c}^{(-)}) = 0
\end{align}
as
\begin{align}
  \theta_{\rm c}^{(-)}
  =
  \arccos
  \left[
    -\frac{2\nu}{\sqrt{h_{\rm k}^{2} - 16\alpha^{2}}}
    \right].
\end{align}
which is valid for $\nu \le 0$ because the convergence value of
$\theta$ should satisfy $0\le\theta_{\rm c}^{(-)}\le \pi/2$.

The convergence value of $\theta$ for the case with
$\lim_{\tau\to\infty}\Delta\phi = \Delta\phi_{\rm c}^{(+)}$ is obtained
in a similar manner as
\begin{align}
  \theta_{\rm c}^{(+)}
  =
  \arccos
  \left[
    \frac{2\nu}{\sqrt{h_{\rm k}^{2} - 16\alpha^{2}}}
    \right],
\end{align}
which is valid for $\nu \ge 0$,  
At $\nu = 0$, $\theta_{\rm c} = \pi/2$ and both
$\Delta\phi_{\rm c}^{(-)}$ and $\Delta\phi_{\rm c}^{(+)}$ are the
valid solutions. Since $\Delta\phi_{\rm c}^{(-)} \neq \Delta\phi_{\rm
  c}^{(+)}$ it might be difficult to achieve a stable oscillation.

The general expressions of the convergence values of $\theta$ and $\Delta\phi$ are summarized as follows
\begin{align}
  \label{eq:fc}
  &
  \Delta\phi_{\rm c}
  = -\frac{\pi(\nu+|\nu|)}{4}+\frac{\nu}{|\nu|}\frac{1}{2}\arcsin\left(\frac{4\alpha}{h_{\rm k}}\right),
  \\
  &
  \label{eq:qc}
  \theta_{\rm c}
  =
  \arccos
  \left[
    \frac{2|\nu|}{\sqrt{h_{\rm k}^{2} - 16\alpha^{2}}}
    \right].
\end{align}
The synchronization region of $\nu$ is obtained by requiring
$\theta_{\rm c}>0$ as
\begin{align}
  \label{eq:sync_region_nu}
  \left| \nu \right| <
  \frac{1}{2}\sqrt{h_{\rm k} - 16 \alpha^{2}}.
\end{align}
The synchronization condition of $h_{\rm k}$ is obtained in a similar
manner as
\begin{align}
  \label{eq:sync_region_hk}
  h_{\rm k} >
  2\sqrt{\nu^{2} + 4\alpha^{2}}.
\end{align}
Taking the limit of $\nu\to 0$ the critical value of $h_{\rm k}$ is
obtained as 4$\alpha$.

%========================================
% Fig. 5
%========================================

\begin{figure}
  \centerline{
    \includegraphics[width=\columnwidth]{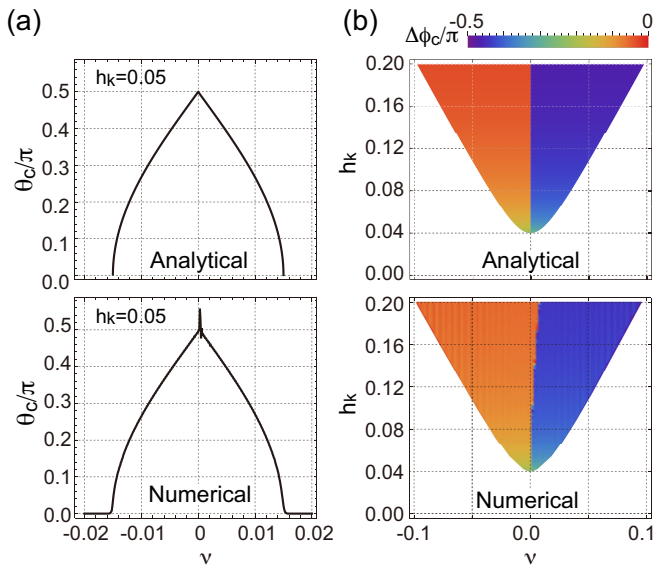}}
\caption{
  \label{fig:sync_region}
  (Color online)
  (a)
  $\nu$ dependence of $\theta_{\rm c}$ obtained by the analytical
  calculation (top) and by the numerical simulation at $\tau = 20000$ (bottom).
  (b)
  A color density plot of $\Delta\phi_{\rm c}$ obtained by the analytical
  calculation (top) and by the numerical simulation at $\tau = 20000$
  (bottom) on the $\nu$-$h_{\rm k}$ plane.
}
\end{figure}

Figure \ref{fig:sync_region}(a) shows $\nu$ dependence of $\theta_{\rm c}$.
The analytical results of Eq. \eqref{eq:qc} is shown in the top
panel. The  numerical results obtained by solving
Eqs. \eqref{eq:LLGtheta} and  \eqref{eq:LLGphi} for $\tau\le$ 20000
are shown in the bottom panel. Although the analytical results are almost
identical to the numerical ones, the small sharp peak appears in the
vicinity of $\nu = 0$ in the numerical results. At this value of $\nu$
the numerical simulations do not converge even at $\tau = 20000$.

Figure \ref{fig:sync_region}(b) is the color density plot of the phase
shift in the synchronization region on the $\nu$-$h_{\rm k}$ plane.
The analytical results of Eqs. \eqref{eq:fc} and
\eqref{eq:sync_region_hk} are shown in the top
panel. The numerical results obtained by solving
Eqs. \eqref{eq:LLGtheta} and  \eqref{eq:LLGphi} for $\tau\le$ 20000
are shown in the bottom panel, where the synchronization region is
defined to satisfy $\theta > $ 0.01 at $\tau = 20000$. 
In both panels the red and blue tones represent $\Delta\phi_{\rm c}^{(-)}$
and $\Delta\phi_{\rm c}^{(+)}$, respectively.
The analytical results are almost identical to the numerical
ones. However the boundary between
 $\Delta\phi_{\rm c}^{(-)}$ and $\Delta\phi_{\rm c}^{(+)}$ in the
numerical results shifts slightly toward larger $\nu$ from the
analytical boundary at $\nu = 0$. This small deviation is caused by the
terms with $\alpha h_{\rm k}$ which we neglect in the analytical
calculations.

%========================================
% Summary
%========================================
\section{Summary}
In summary, the effects of microwave voltage on the magnetization
dynamics in the FL of a VCMA-based MRAM is theoretically analyzed.
It is shown that the large angle precession of magnetization
is maintained if the angular frequency of the microwave voltage is
about twice the natural angular frequency of the precession.
The effective equations of motion for the slow dynamics of the polar
angle and the phase shift are derived.
The mechanism of the synchronization is explained by analyzing the
slow dynamics of the polar angle, phase shift, and effective
damping coefficient. The phase shift is automatically adjusted to
eliminate the effective damping. The convergence value of the phase
shift strongly depends on the sign of the angular frequency detuning.
The synchronization conditions of the angular frequency detuning and
the amplitude of the oscillating anisotropy field are obtained.
The critical value of the amplitude of the oscillating anisotropy
field is proportional to the Gilbert damping constant.
The results are useful for development of the VCMA-based
energy-efficient spintronics devices using magnetization precession
such as a VCMA-based MRAM and a nano-scale microwave magnetic field
generator.

%========================================
% Acknowledgement
%========================================
\acknowledgements
We acknowledge T. Nozaki and T. Yamamoto for useful discussions.
This work was partly supported by JSPS KAKENHI Grant Numbers JP19K05259
and JP19H01108.

%========================================
% References
%========================================
%% \bibliography{MW_VCMA_Excitation.bib}


\end{document}